\documentclass[
]{ceurart}

\usepackage{listings}
\usepackage{pgfplots}
\usepackage{pgfplotstable}
\begin{document}

\copyrightyear{2025}
\copyrightclause{Copyright for this paper by its authors.
  Use permitted under Creative Commons License Attribution 4.0
  International (CC BY 4.0).}

\conference{First Edition of COL-SAI Workshop "COllaboration and Learning through Symbiotic Artificial Intelligence" co-located CHItaly 2025 - Salerno, Italy}

\title{Leveraging Trustworthy AI for Automotive Security in Multi-Domain Operations: Towards a Responsive Human-AI Multi-Domain Task Force for Cyber Social Security}

\tnotemark[1]
\tnotetext[1]{You can use this document as the template for preparing your
  publication. We recommend using the latest version of the ceurart style.}

\author[1]{Vita Santa Barletta}[%
orcid=0000-0002-0163-6786,
email=vita.barletta@uniba.it]
\fnmark[1]
\address[1]{Università degli studi di Bari Aldo Moro, Piazza Umberto I, 70121 Bari, Apulia, Italy}

\author[1]{Danilo Caivano}[%
orcid=0000-0001-5719-7447,
email=danilo.caivano@uniba.it]
\fnmark[1]

\author[1]{Gabriel Cellammare}[%
email=g.cellammare1@studenti.uniba.it]
\fnmark[1]

\author[2]{Samuele del Vescovo}[%
orcid=0009-0001-8220-5135,
email=samuele.delvescovo@imtlucca.it]
\cormark[1]
\fnmark[1]
\address[2]{Scuola IMT Alti Studi Lucca, Piazza S.Francesco, 19, 55100 Lucca, Italy}

\author[1]{Annita Larissa Sciacovelli}[%
orcid=0000-0003-4795-9238,
email=annitalarissa.sciacovelli@uniba.it]
\fnmark[1]
\cortext[1]{Corresponding author.}
\fntext[1]{These authors contributed equally.}

\begin{abstract}
Multi-Domain Operations (MDOs) emphasize cross-domain defense against complex and synergistic threats, with civilian infrastructures like smart cities and Connected Autonomous Vehicles (CAVs) emerging as primary targets. As dual-use assets, CAVs are vulnerable to Multi-Surface Threats (MSTs), particularly from Adversarial Machine Learning (AML) which can simultaneously compromise multiple in-vehicle ML systems (e.g., Intrusion Detection Systems, Traffic Sign Recognition Systems). Therefore, this study investigates how key hyperparameters in Decision Tree-based ensemble models—Random Forest (RF), Gradient Boosting (GB), and Extreme Gradient Boosting (XGB)—affect the time required for a Black-Box AML attack i.e. Zeroth Order Optimization (ZOO). Findings show that parameters like the number of trees or boosting rounds significantly influence attack execution time, with RF and GB being more sensitive than XGB. Adversarial Training (AT) time is also analyzed to assess the attacker's window of opportunity. By optimizing hyperparameters, this research supports Defensive Trustworthy AI (D-TAI) practices within MST scenarios and contributes to the development of resilient ML systems for civilian and military domains, aligned with Cyber Social Security framework in MDOs and Human-AI Multi-Domain Task Forces.
 
\end{abstract}

\maketitle

\section{Introduction} \label{sec:intro}

Multi-Domain Operations (MDOs) constitute the central paradigm of modern military strategy, emphasizing the integrated coordination of capabilities across \textit{Space}, \textit{Air}, \textit{Land}, \textit{Sea}, and \textit{Cyber} domains to produce synergistic malicious effects \cite{mdo_economic}. These operations generate complex, multi-surfaced challenges aimed at complicate the adversary’s Military Decision-Making Process (MDMP) \cite{mdo_future_nato, mdo_oe_1}. Within this framework, Multi-Domain Task Forces (MDTFs) play a key role by aligning defence resources across both physical and informational environments \cite{mdtf}. A core challenge in MDOs is the lack of defensive synchronization between kinetic and cyber capabilities across domains \cite{mdo_capabilities}. To address this, the Cyber Social Security (CSS) framework has been introduced to integrate cyber defence within traditional domain-based strategies, forming the CSS-MDO framework \cite{cyber_social_security, cyber_social_security_frame}. 

The CSS-MDO framework adopts a multidimensional approach, where the five active warfare domains are represented along the horizontal axis, each equipped with specialized tools, methods, and procedures \cite{cyber_social_security_frame}. These are aligned with the vertical axis representing the Detection-Response-Prevention processes. In this framework, unlike other military domains, the "cyber" domain is not only a "link" between the various "traditional" domains but it has its own offensive/defensive identity \cite{mdo_oe_1}. This bi-dimensional structure identify operational tiers (i.e. horizontally and vertically) in which MDTFs may operate \cite{ida}.

Smart cities constitute vulnerable assets within future MDOs, especially involving the \textit{Cyber} and \textit{Land} domains. Among the critical components of a smart city's Internet of Things (IoT) infrastructure, Connected and Autonomous Vehicles (CAVs) represent a prominent point of vulnerability \cite{smart_city_auto}. As foundational enablers of shared and electric mobility, CAVs are central to optimizing urban transportation and improving sustainable travel models \cite{smart_transp, automotive_tech}. However, in this rapidly evolving and innovation context, this high-value assets' attack surfaces are expanding and complicating in ways challenging the capacity of any "cyber-social" blue team to effectively detect and mitigate threats. Specifically, attackers (more or less skilled) may exploit known vulnerabilities in In-Vehicle Networks (IVNs) based on the Controller Area Network (CAN) protocol \cite{stato_ids_1_5}. These attacks may be part of broader, coordinated military/civilian activities conducted across multiple domains to create complex and orchestrated dilemmas for defenders \cite{Tommasi_2019}.

Within the context of MDOs and the realted Multi-Domain Threats (MDTs), one of the attackers goals may be to compromise Electronic Control Units (ECUs), which function as critical communicational nodes within IVNs \cite{stato_ids_1}. This threat can also affect the  autonomous vehicle's driving system by exploiting various IVNs levels. Such intrusions can lead to anomalous or unsafe behavior in the affected vehicle, thereby posing direct risks to its functionality and safety \cite{auto12}. Furthermore, an attacker can compromise physically road signs to compromise autonomous driving's features of the vehicle. In response to these threats, the integration of Artificial Intelligence (AI) and Machine Learning (ML) technologies has emerged as a promising avenue for reinforcing vehicular cybersecurity. Among these approaches, there are ML-based Intrusion Detection Systems (IDSs) as well as Traffic Sign Recognition Systems (TSRS), which are specifically designed to detect abnormal communication patterns and unauthorized access attempts within IVNs \cite{stato_ids_1_9}.

So, 
an adversary can manipulate input data, such as images or CAN bus frames, at the testing or deployment phase \cite{enisa_securing_ai}. These perturbations are imperceptible to human observers yet are sufficient to deceive the targeted ML model into producing incorrect classifications \cite{stato_aml_6, stato_aml_3}. AML attacks are generally categorized into three scenarios, discriminated by the level of knowledge the attacker possesses about the internal architecture and parameters of the victim model. Among these, the Black-Box setting is considered the most plausible and accessible from an adversarial perspective, as it assumes no prior access to or knowledge of the model’s internal workings \cite{aml_1, aml_2}. Moreover, aligning with the conceptual framework of MDTs, in which adversaries operate across multiple domains simultaneously, it is plausible to hypothesize scenarios in which attacks are exploited on different surfaces of a single asset. This concept can be referred to as Multi-Surface Threat (MST). Current literature on the application or conceptualization of Black-Box attacks within the CAN bus frame detection task remains limited and in an early stage of development (even in MDTs and MSTs scenarios).

Therefore, this paper has the primary goal of investigate about the role of specific hyperparameters associated with Decision Tree (DT)-based ensemble Technology Transfer (TT) models i.e. Random Forest (RF), Gradient Boosting (GB), and Extreme Gradient Boosting (XGB). These are the core of the supervised ML-based victim IDS for the CAN bus frame detection task. It is assumed that the IDS (installed onboard the vehicle) is subjected to a Black-Box AML attack i.e. the Zeroth Order Optimization (ZOO) in a pure evasive Black-Box setting. This type of attack is conceptualized as a part of a MST, so a Single-Surface Threat (SST), falling into the \textit{Cyber} component of a complex MDT during a MDO. So, this work tries to address how variations in selected hyperparameters affect the time required to generate adversarial examples for each targeted ML model. Results indicate that the number of bagging trees in RF and the number of boosting rounds in GB have a significant impact on the attack time. Thus, the same does not hold for the boosting rounds in XGB. These hyperparameters, in the cases of RF and GB, can be interpreted as intrinsic (or deterrence) defense against the ZOO attack. Appropriately values for these hyperparameters may lead to a trustworthy AI-by-design approach for In-Vehicles (I-V) ML systems' robustness \cite{enisa_ai_cyber}, \cite{enisa_securing_ai} contributing to the concept of Defensive Trustworthy AI (D-TAI) i.e. AI for defence purposes. In particular, the work underscores the relevance of robustness \cite{robustness} and security \cite{security_nist} properties in the design and deployment of ML models within adversarial environments \cite{resp_ai_2}. Additionally, Adversarial Training (AT) time is analyzed to better understand the attacker’s "window of opportunity", proving the combo between hyperparameters' correct tuning and AT can slow the attacker's steps, fostering the activities of the blue team. Finally, this work aims to contribute to the proper education regarding responsible development of ML systems in both civilian and military (public/private) contexts for industries and academies, promoting their integration into the CSS-MDO \cite{cyber_social_security_frame} framework as part of a defence strategy coordinated by Human-AI Multi-Domain Task Forces (H-AI MDTF). The proof of concept is the qualitative identification of the positive impact of this trustworthy AI-by-design practice on the vertical axes of the CSS-MDO framework.

In summary, the research questions (RQs) are:
\begin{itemize}
    \item RQ1: Can the hyperparameters related to the number of bagging trees in RF, the number of boosting rounds in GB, and the number of boosting rounds in XGB affect the time needed to generate ZOO adversarial examples, when applied to supervised ML-based IDS for the CAN bus frame detection task (in a Black-Box attack scenario)?
    \item RQ2: Can the combination of the timing for the AT and the correct setting of the previous parameters slow down the attacker's actions by reducing the attacker's "window of opportunity"?
    \item RQ3: Is it possible to qualitatively quantify the (positive) impact of these values on the Detection-Response-Prevention axes of the CSS-MDO framework (for H-AI MDTFs educational purpose)?
\end{itemize}

So, the main contributions are:
\begin{itemize}
    \item Empirically evaluate the influence of hyperparameters as the number of bagging trees in RF, the number of boosting rounds in GB, and the number of boosting rounds in XGB on the time needed to generate adversarial examples exploiting the ZOO attack, when applied to supervised ML-based IDS in the CAN bus frame detection task (considering a Black-Box threat model).
    \item Understand whether the combination of the correct setting of the previously mentioned parameters and the time needed to execute the AT can slow down or reduce the attacker's "window of opportunity".
    \item Qualitatively quantify the (positive) impact of these values on the "Detection", "Response" and "Prevention" axes of the CSS-MDO framework (for H-AI MDTFs educational purpose).
\end{itemize}

The paper is organized as follows: \autoref{sect:rel} describes the related works; \autoref{sect:meth} describes the experimentation setup; \autoref{sect:resu} shows the results; \autoref{sect:conc} conclude the work and explain the future developments.

\section{Related Work \& Research Gap} \label{sect:rel}

\subsection{Evasion Black-Box AML for CAN Bus Frame Detection}
\label{sect:related_work.1}
To the best of our knowledge, the scientific literature addressing Black-Box AML attacks against ML-based IDS in the context of CAN bus frame detection remains relatively underdeveloped actually.

Zenden et al. \cite{stato_aml_2} examined the Fast Gradient Sign Method (FGSM) attack's impact on various ML and DL models' performance within a surrogate Black-Box attack scenario. Their study further demonstrated that adversarial training serves as an effective mitigation strategy, yielding notably positive outcomes. The evaluated models included BL-DNN, BL-Ensemble, SOTA-CNN, and SOTA-LSTM architectures \cite{stato_aml_2}. The experiments were conducted using a subset of the Survival dataset \cite{survival_dataset}. The results indicated that ML models were particularly susceptible to the FGSM attack, exhibiting an accuracy degradation of approximately 40\%.

Longari et al. \cite{stato_aml_3} proposed a novel methodology for executing Black-Box evasion attacks (in a pure scenario) against ML-based IDS within the context of online CAN bus frame detection. Their method targets the entire transmission flow by analyzing segments of CAN payloads. The attack strategy employs a sliding window technique over the payload data, rather than processing entire examples in isolation \cite{stato_aml_3}. The experimentation utilizes the “ReCAN” dataset, specifically the “ID C-1” subset collected from a real Alfa Romeo Giulia Veloce vehicle \cite{stato_aml_3}. A range of ML algorithms were evaluated as the IDS core, including Small-LSTM, Small-GRU, Large-GRU, CANnolo, Neural Network, Vector Auto-Regressive (VAR), and Hamming distance-based models \cite{stato_aml_3}.

Instead, Aloraini et al. \cite{stato_aml_6} have conducted an adversarial attack using a substitute victim IDS, trained on data extracted from the OBD-II interface. This dataset is different from the one used to train the real victim IDS \cite{stato_aml_6}. This scenario constitutes a non-pure Black-Box due to the transferability of the adversarial examples exploited \cite{stato_aml_6}. The victim IDS models were: a baseline proprietary IDS based on Deep Neural Network (DNN) and one state-of-the-art model, i.e. MTH-IDS. The surrogated models were a DNN and a DT. The dataset exploited for the surrogated model is the Car Hacking Dataset \cite{car_hacking_dataset}. Several White-Box AML attacks were considered like FGSM, Basic Iterative Method (BIM), Projected Gradient Descent (PGD) and Jacobian-based Saliency Map Attack (JSMA) \cite{stato_aml_6}. The experimental results have shown the decrease of the F1 scores from 95\% to 38\% and from 97\% to 79\% respectively for the real victim models \cite{stato_aml_6}.  

These works do not consider attacks conducted directly against the target IDS in a pure Black-Box scenario in a TT context. Moreover, they do not explore the application of state-of-the-art Black-Box AML techniques that are not explicitly tailored to this specific task. In contrast, the study by Barletta et al. \cite{delVescovo_TechDefense_2024} investigates the application of the ZOO attack within a pure Black-Box setting for the same task, focusing on supervised ML algorithms. Originally developed for image recognition tasks, the ZOO attack was exploited using the OTIDS dataset \cite{OTIDS_Dataset_2017}. The victim models included DT, RF, GB, and XGB (contestualized in TT scenarios). Experimental findings revealed a reduction in weighted accuracy of approximately 70\%. Furthermore, adversarial training was once again validated as an effective countermeasure against such attacks. 

 \subsection{Evasion Black-Box AML Attacks \& Defensive Trustworthy AI in CAN Bus Frame Detection for MDOs}
\label{sect:related_work.2}

This paper aims to explore the concept of trustworthy AI for defensive purposes (specifically the robustness dimension) as a proactive defense mechanism (i.e. D-TAI) for ML-based systems in the automotive domain. Among existing countermeasures, AT is the most widely adopted technique to enhance the security posture of ML-based systems and mitigate the risks posed by Black-Box AML attacks \cite{riferimento_2, survey_robust_ai, delVescovo_TechDefense_2024}. However, the current literature reveals a significant gap concerning the role of robustness-by-design oriented programming practices in the development of ML-based systems for CAN bus frame detection under Black-Box AML attack scenarios.

\subsection{Defensive Trustworthy AI Impact Evaluation on CSS-MDO Framework for H-AI MDTFs}

It is necessary to consider the impact of best ML-based systems programming practices (for D-TAI and especially for robustness of ML models) on Single Surface threats (considered as a part os MSTs targetting various I-V Systems) mapped within the CSS-MDO framework's policies. In other words, it is necessary to consider the positive impact of properly educating MDTFs about these practices emphasizing the virtuous collaboration between human and (well-setted) artificial agents in H-AI MDTFs. All of that can be a piece of the MDTs big puzzle in MDOs scenario. Accordingly, this paper seeks to underscore this critical need for future research.

\label{sect:related_work.3}

\section{Methodology} \label{sect:meth}

In this section (useful for answering all RQs), details about the Black-Box attack scenario, the ZOO attack pipeline, the empirical estimation of attack and AT time and the qualitative analysis of the parameters' impact on the CSS-MDO framework's axes are discussed. All this is about the CAN-based single surface threat. This work is based on Python 3.9. The implementation of the ML models is provided by the Scikit-learn library. The XGBoost model implementation is based on the xgboost library. The Pandas framework is used to handle the dataset. The ZOO attack implementation is provided by the Adversarial Robustness Toolkit (ART) \cite{art}. The working machine is supported by an AMD Ryzen 5 2600 Six-Core Processor and 16 GB of RAM. 

\subsection{CAN-based SST Attack Scenario}
\label{sect:meth.1}
\begin{figure}
    \centering
    \includegraphics[width=0.7\linewidth]{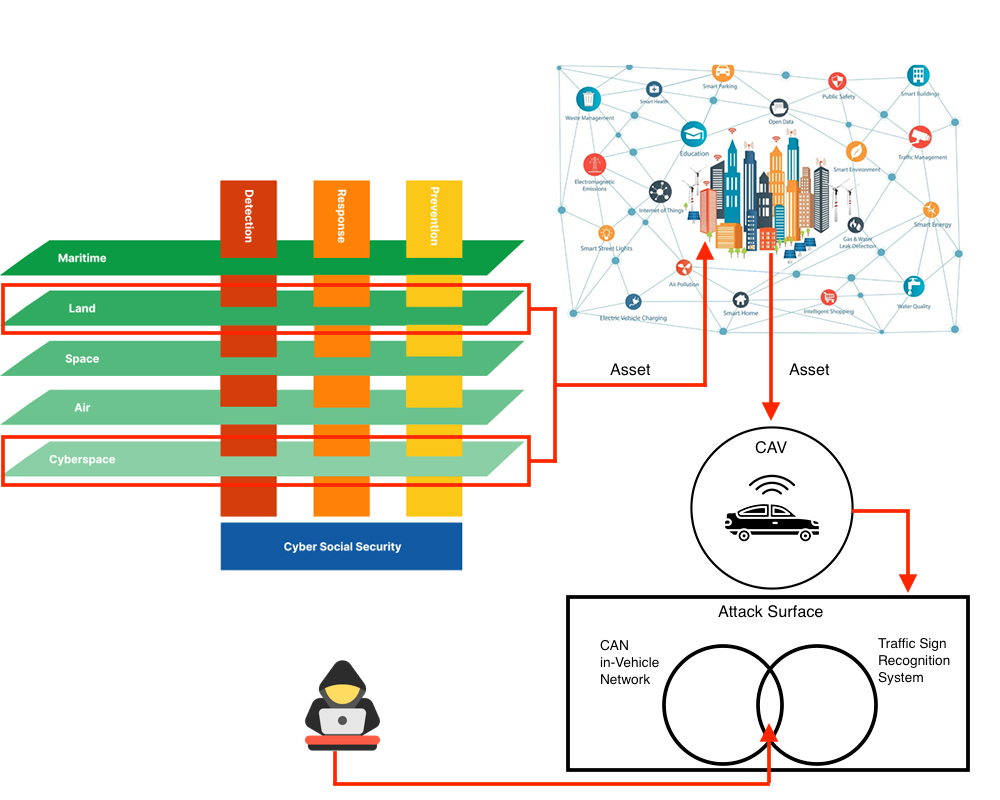}
    \caption{MST considered}
    \label{fig:meth}
\end{figure}

This phase addresses the RQ1. Considering the possibility of MSTs (in the examined case impacting several I-V systems simultaneously), an example of that could be a simultaneous threat on the CAN-based IVN protected by an IDS for the CAN bus frame detection task and on the network enabling the Traffic Sign Recognition System (TSRS). Figure \ref{fig:meth} describe the attack scenario. 

The CAN-based SST attack begins with a Vulnerability Assessment and Penetration Testing (VAPT) phase targeting the CAN based In-Vehicle Network, aiming to compromise a single ECU. This initial step facilitates both the exfiltration and injection of CAN frames, enabling the attacker to infer behavioral patterns of the target IDS. The ultimate goal is to penetrate the IDS module itself \cite{tech_transfer}, thereby obtaining the true label assigned to each preprocessed frame for the generation of corresponding adversarial examples. Additionally, the attacker may observe the IDS’s output by compromising a module that interfaces with the IDS. Importantly, the attacker operates under a pure Black-Box scenario, with no prior knowledge of the victim system's internal architecture or parameters \cite{delVescovo_TechDefense_2024}.

\subsection{Attack Pipeline for the CAN-based SST}
\label{sect:meth.2}
This work is based on the Barletta et al. \cite{delVescovo_TechDefense_2024} attack pipeline, useful for training the victim ML models. The OTIDS dataset \cite{OTIDS_Dataset_2017} is prelaborated following the Bari et al. \cite{riferimento_2} pipeline.  
The final dataset version is splitted into three parts: \(A\) (i.e. the 60\% part), \(B\) and \(C\) (i.e. the other 20\% parts). ML models useful for empirical estimation are: RF bagging-based, GB and XGB (in their default configurations). The attack pipeline extracted is composed by these steps (for each victim ML model):
\begin{enumerate}
    \item IDS training on the \(A\) dataset (obtaining \(Model_A\));
    \item Adversarial examples sets' generation i.e. \(B'\) and \(C'\) (starting from the \(B\) and \(C\) sets) on \(Model_A\);
    \item Training on \(A+B+B'\) dataset, obtaining \(Model_{A+B+B'}\) (Adversarial Training);
    
\end{enumerate}

A K-Fold Stratified Cross Validation (K-FSCV) (with \(k=5\)) is performed before running the ZOO attack on the \(A\) subdataset. The ZOO attack follows the default configuration except for:
\begin{itemize}
    \item the \textit{learning\_rate} setted to 0.1 (default is 0.01). The attacker probably want to converge very fast (during the gradient descent);
    \item the \textit{max\_iter} setted to 50 (default is 10). The attacker probably want to get examples very close to the normal ones (by increasing the number of trials);
    \item the \textit{variable\_h} setted to 0.2 (default is 0.0001). The attacker probably wants the adversarial examples quickly (enlarging the extremes of the search range); 
\end{itemize}

This research work is not interested in quantifying the impact of the attack on the victim ML performances' since Barletta et al. \cite{delVescovo_TechDefense_2024} have already explored that. For this reason, phase three of the attack pipeline directly considers the execution of the adversarial training.

\subsection{Empirical Estimation of the Hyperparameters' Influence \& AT time on the CAN-based SST}
\label{sect:meth.3}

This subsection is useful for answering RQ1 and RQ2. Ideally, a Vehicle-Security Operations Center (Vehicle-SOC) involved in a MDTF would prefer to exploit a ML-based IDS that maximizes the time required for generating adversarial examples while minimizes the coutermeasure's time, thereby increasing the attack’s operational cost. Considering that, certain hyperparameters of ensemble-based ML models specifically RF, GB, and XGB can be conceptualized as intrinsic defensive mechanisms, potentially influencing the computational effort needed to generate input adversarial examples.
The approach adopted in this study involves measuring the time (seconds) required to generate 92270 adversarial examples for each model (i.e. RF, GB, XGB), as a function of incrementally varied hyperparameter values. Time is detected after about five minutes of computation (for each involved ML model and hyperparameter value). These include the number of bagging trees in RF and the number of boosting rounds in both GB and XGB. The empirical assessment is conducted on \(Model_A\), focusing on the second phase of the previously described attack pipeline, and it is limited to the \(B'\) dataset, assuming its consistent with \(C'\). 
The goal is to determine whether a direct proportional relationship exists between these hyperparameters and the adversarial example generation time. This analysis seeks to provide actionable insights into optimal hyperparameter configurations and the selection of the most defensive ensemble model. Finally, only for models demonstrating a "defensive" tendency of the mentioned hyperparameters is the time required to perform an AT (considering the training set \(A+B+B'\) consisting of 461350 examples) evaluated as the values of the hyperparameters vary. This is useful to better understand the extent of the attacker's "window of opportunity." It is assumed that once introduced into the CAN network, the attacker proceeds with the compromise of the IDS system.

\subsection{Hyperparameters' Impact Qualitative Analysis on CSS-MDO Framework}
\label{sect:meth.4}
This subsection is useful for answering RQ3. Considering the critical nature of the scenario that surrounds this research work, an impact (positive) analysis realted to the MDTFs education (acting along the CSS-MDO framework vertical axes) about the best programming practices that improve the resilience of ML models to Black-Box attacks is an important milestone to underline the right importance of these. All this is intended to emphasize the fruitful collaboration between human and artificial agents in future Human-AI MDTFs. The actual qualitative analysis is based on the \textit{Land} and \textit{Cyber} domains. This analysis comes from an high-level qualitative risk assessment related to this SST. This second analysis is adopted considering multiple hypothetical negative consequences: the potential to incite a climate of terror through anomalous vehicle behavior and the cognitive disruption of civilian and military operators, the reputational demage to the national infrastructure and institutions \cite{delVescovo_TechDefense_2024}, and the inherently risk-averse perspective guiding the human evaluator point-of-view \cite{Baldassarre2019}.

\section{Result \& Discussion} \label{sect:resu}

\subsection{Empirical Evalution of Attack \& AT Time}
\label{sect:resu.1}

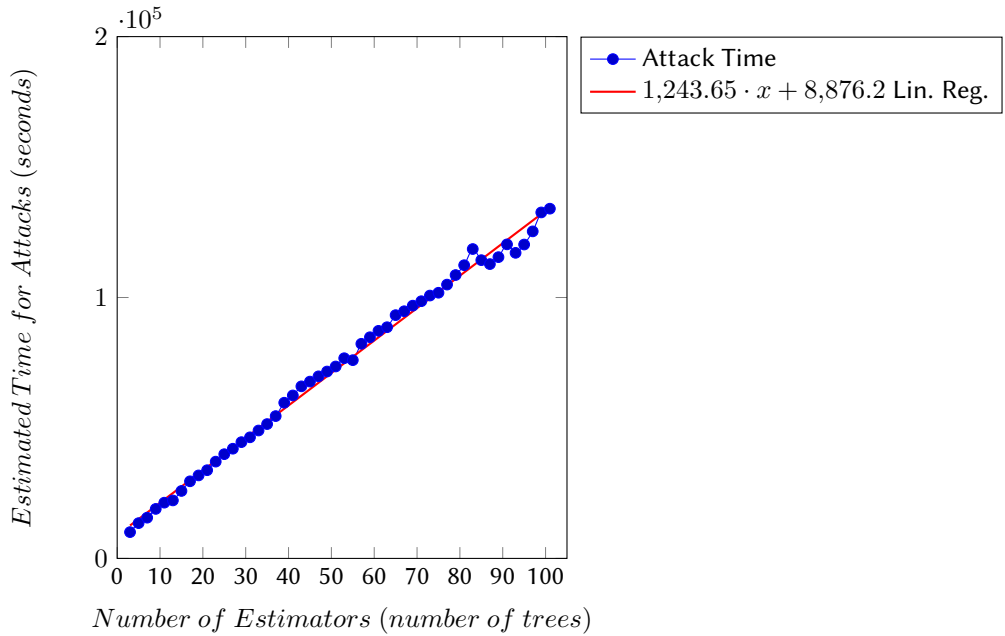
\begin{figure}
\begin{centering}
    \begin{tikzpicture}
\begin{axis}[
    xlabel=$Number\ of\ Estimators\ (number\ of\ trees)$,
    ylabel=$Estimated\ Time\ for\ Attacks\ (seconds)$,
    xmin=0, xmax=105,
    ymin=0, ymax=200000,
    xtick={0,10,...,105},
    xticklabels={0,10,...,110},   
    ytick={0,100000,...,900000},
    width=7.5cm,
  	height=8.5cm,
    legend pos=outer north east,
    legend cell align=left,
    smooth,
            ]
\addplot table [x=trees, y=time, col sep=comma] {./csv/est_zoo_rf.csv};
\addlegendentry{Attack Time}

\addplot [no markers, thick, red] table [x=trees, y={create col/linear regression={y=time, x=trees}}, col sep=comma] {./csv/est_zoo_rf.csv};

\addlegendentry{%
        $\pgfmathprintnumber{\pgfplotstableregressiona} \cdot x
        \pgfmathprintnumber[print sign]{\pgfplotstableregressionb}$ Lin. Reg.}%
\end{axis}
    \end{tikzpicture}
\caption{Time required to generate \(B'\) as a function of the increasing number of bagging trees in the RF model.}
\label{fig:estimated_rf}
\end{centering}
\end{figure}

Figure \ref{fig:estimated_rf} presents an estimation of the time required to generate 92270 adversarial examples (corresponding to subset \(B'\)) as a function of the number of bagging trees (i.e. the examinated hyperparameter) in the RF model. The results demonstrate a clear linear relationship between the number of estimators and the generation time, as evidenced by the regression line. For instance, with 81 estimators, the expected generation time approaches approximately 31 hours. This finding indicates that the selected hyperparameter influences not only the model’s predictive performance but also its robustness against ZOO attack. Therefore, the response to RQ1, concerning the RF model, is affirmative.

\begin{figure}
\begin{centering}
    \begin{tikzpicture}
\begin{axis}[
    xlabel=$Number\ of\ Estimators\ (number\ of\ boosting\ rounds)$,
    ylabel=$Estimated\ Time\ for\ Attacks\ (seconds)$,
    xmin=0, xmax=110,
    ymin=0, ymax=20000,
    xtick={0,10,...,110},
    xticklabels={0,10,...,110},   
    ytick={0,10000,...,50000},
    width=7.5cm,
  	height=8.5cm,
    legend pos=outer north east,
    legend cell align=left,
    smooth,
            ]

\addplot table [x=boost, y=time, col sep=comma] {./csv/est_zoo_gb.csv};
\addlegendentry{Attack Time}

\addplot [no markers, thick, red] table [x=boost, y={create col/linear regression={y=time, x=boost}}, col sep=comma] {./csv/est_zoo_gb.csv};

\addlegendentry{%
        $\pgfmathprintnumber{\pgfplotstableregressiona} \cdot x
        \pgfmathprintnumber[print sign]{\pgfplotstableregressionb}$ Lin. Reg.}%
\end{axis}

    \end{tikzpicture}
\caption{Time required to generate \(B'\) as a function of the increasing number of boosting rounds in the GB model.}
\label{fig:estimated_gb}
\end{centering}
\end{figure}
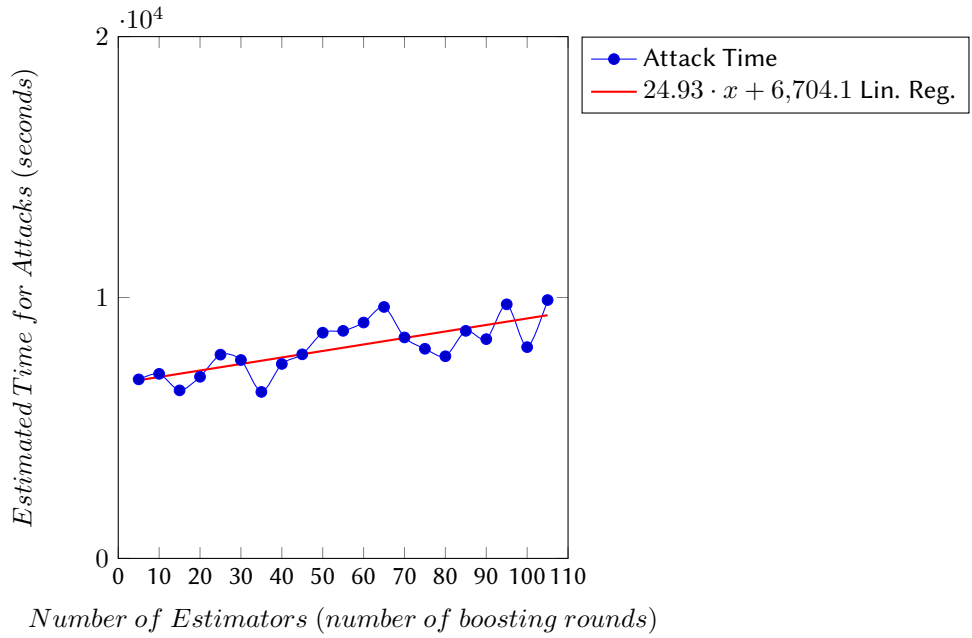

\begin{figure}
\begin{centering}
    \begin{tikzpicture}
\begin{axis}[
    xlabel=$Number\ of\ Estimators\ (number\ of\ boosting\ rounds)$,
    ylabel=$Estimated\ Time\ for\ Attacks\ (seconds)$,
    xmin=0, xmax=110,
    ymin=0, ymax=100000,
    xtick={0,10,...,110},
    xticklabels={0,10,...,110},   
    ytick={0,10000,...,200000},
    width=7.5cm,
  	height=8.5cm,
    legend pos=outer north east,
    legend cell align=left,
    smooth,
            ]

\addplot table [x=boost, y=time, col sep=comma] {./csv/est_zoo_xgb.csv};
\addlegendentry{Attack Time}

\addplot [no markers, thick, red] table [x=boost, y={create col/linear regression={y=time, x=boost}}, col sep=comma] {./csv/est_zoo_xgb.csv};

\addlegendentry{%
        $\pgfmathprintnumber{\pgfplotstableregressiona} \cdot x
        \pgfmathprintnumber[print sign]{\pgfplotstableregressionb}$ Lin. Reg.}%
\end{axis}

    \end{tikzpicture}
\caption{Time required to generate \(B'\) as a function of the increasing number of boosting rounds in the XGB model.}
\label{fig:estimated_xgb}
\end{centering}
\end{figure}
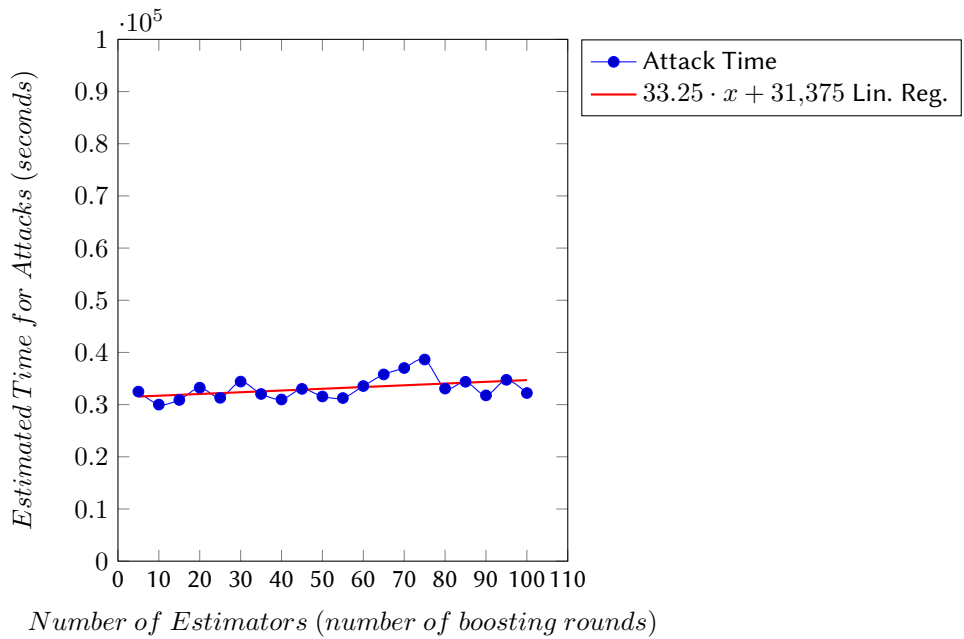

Figure \ref{fig:estimated_gb} presents the same type of the previous estimation but applied to the GB model. In this case, the examined independent variable is the number of boosting rounds. The results clearly reveal a directly proportional relationship between the number of boosting rounds and the time required to generate the adversarial examples, as indicated by the regression line. For instance, approximately 2 hours are needed for 80 estimators. These findings support the same conclusions drawn for the RF model, thereby affirmatively answering RQ1 in the context of GB.

Figure \ref{fig:estimated_xgb} depicts the estimation for the XGB model, again using the number of boosting rounds as the independent variable. Unlike the RF and GB cases, the results do not exhibit a consistent linear relationship between the number of estimators and the adversarial generation time, as confirmed by the regression analysis. For example, around 9 hours are required when 80 estimators are used. Consequently, the answer to RQ1 in the case of XGB is negative.

\begin{figure}
\begin{centering}
    \begin{tikzpicture}
\begin{axis}[
    xlabel=$Number\ of\ Estimators\ (number\ of\ trees)$,
    ylabel=$Estimated\ Time\ for\ Training\ (seconds)$,
    xmin=0, xmax=105,
    ymin=0, ymax=170,
    xtick={0,10,...,105},
    xticklabels={0,10,...,110},   
    ytick={0,20,...,170},
    width=7.5cm,
  	height=8.5cm,
    legend pos=outer north east,
    legend cell align=left,
    smooth,
            ]
\addplot table [x=trees, y=time, col sep=comma] {./csv/est_time_adv_training_zoo_rf.csv};
\addlegendentry{Training Time}

\addplot [no markers, thick, red] table [x=trees, y={create col/linear regression={y=time, x=trees}}, col sep=comma] {./csv/est_time_adv_training_zoo_rf.csv};

\addlegendentry{%
        $\pgfmathprintnumber{\pgfplotstableregressiona} \cdot x
        \pgfmathprintnumber[print sign]{\pgfplotstableregressionb}$ Lin. Reg.}%
\end{axis}
    \end{tikzpicture}
\caption{Time required to run the AT as a function of the increasing number of bagging trees in the RF model}
\label{fig:estimated_adv_rf}
\end{centering}
\end{figure}
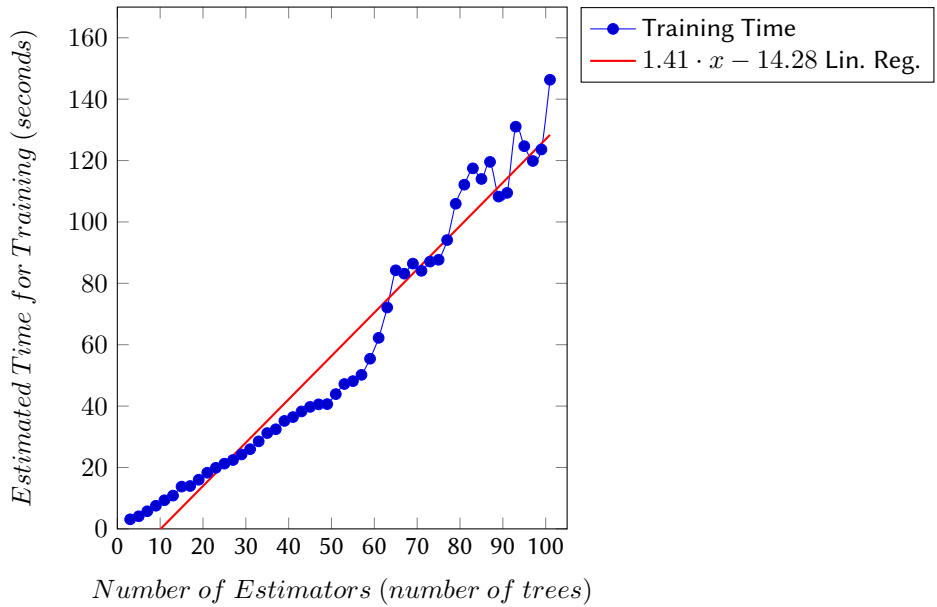

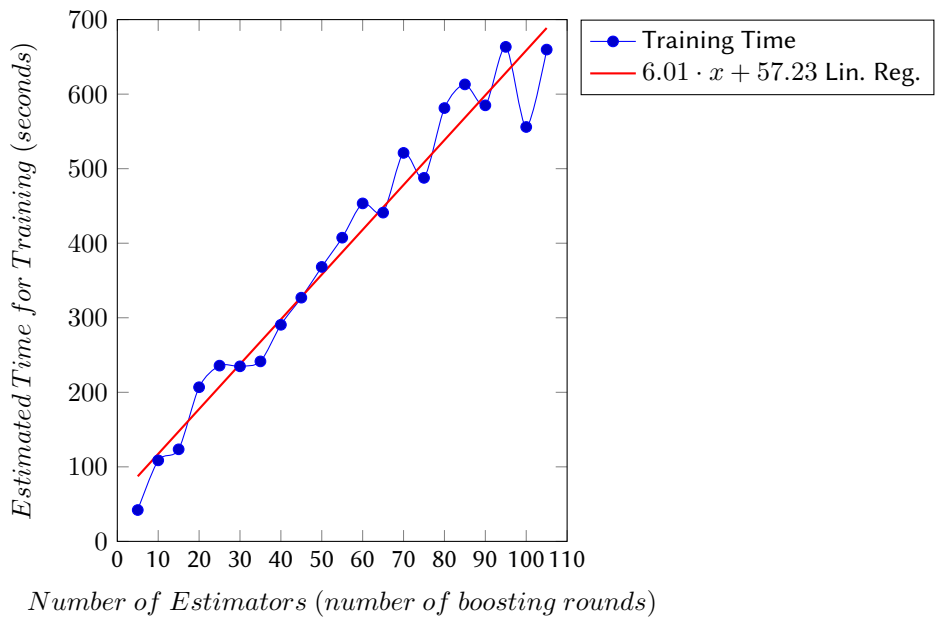
\begin{figure}
\begin{centering}
    \begin{tikzpicture}
\begin{axis}[
    xlabel=$Number\ of\ Estimators\ (number\ of\ boosting\ rounds)$,
    ylabel=$Estimated\ Time\ for\ Training\ (seconds)$,
    xmin=0, xmax=110,
    ymin=0, ymax=700,
    xtick={0,10,...,110},
    xticklabels={0,10,...,110},   
    ytick={0,100,...,700},
    width=7.5cm,
  	height=8.5cm,
    legend pos=outer north east,
    legend cell align=left,
    smooth,
            ]
\addplot table [x=boost, y=time, col sep=comma] {./csv/est_time_adv_training_zoo_gb.csv};
\addlegendentry{Training Time}

\addplot [no markers, thick, red] table [x=boost, y={create col/linear regression={y=time, x=boost}}, col sep=comma] {./csv/est_time_adv_training_zoo_gb.csv};

\addlegendentry{%
        $\pgfmathprintnumber{\pgfplotstableregressiona} \cdot x
        \pgfmathprintnumber[print sign]{\pgfplotstableregressionb}$ Lin. Reg.}%
\end{axis}
    \end{tikzpicture}
\caption{Time required to run the AT as a function of the increasing number of boosting rounds in the GB model}
\label{fig:estimated_adv_gb}
\end{centering}
\end{figure}

Figure \ref{fig:estimated_adv_rf} illustrates the evolution of AT time as a function of the number of bagging trees in the RF model. A clear direct proportionality is observed between the two variables, with a maximum of approximately 150 seconds for 105 trees. Notably, some configurations show local minima in training time, which can be attributed to tree configurations that limit the depth of the learned patterns. A similar phenomenon is observable in Figure \ref{fig:estimated_adv_gb}, which depicts the impact of the number of boosting rounds on training time for the GB model. In this case, the maximum time reaches approximately 670 seconds for 105 boosting rounds. When comparing the estimated attack and AT times for both models, the RF model demonstrates a more favorable trade-off. 

Furthermore, in both scenarios, this combination of training and attack times effectively reduces the attacker’s "window of opportunity" (thus strengthening the defence) during an attack on the In-Vehicle Network (IVN), particularly when considering the potential additional attack delay introduced by the execution of AT. The needed AT time is significantly less than the time provided to the H-AI MDTF to detect the intrusion in the CAN network, making this countermeasure almost necessary in online systems. So, the answer to RQ2 is affirmative.

\subsection{Hyperparameters' Impact on Educating MDTF along CSS-MDO Framework Axes for MDOs}
\label{sect:resu.2}

Table \ref{tab:impact} shows the impact of the previous analysis on the CSS-MDO framework axes for MDOs. This is the answer to RQ3. Generally, the education of MDTFs (even H-AI) about this analysis has a "Very High" impact due to the complementary motivations with respect to \cite{delVescovo_TechDefense_2024}). An "High" impact is observed on the "Prevention" axis since deterrence does not categorically prevent the execution of the attack.

\begin{table*}
\caption{Appropriate values' impact assigned to the hyperameters under consideration for RF and GB on the CSS-MDO framework's axes for MDOs.}
\label{tab:impact}
\begin{centering}
\resizebox{0.6\textwidth}{!}{%
\begin{tabular}{ccc}
\toprule
\textbf{\begin{tabular}[c]{@{}c@{}}Axis of \\ CSS-MDO Framework\end{tabular}} & \textbf{Impact} & \textbf{Motivation} \\ \midrule
Detection & \cellcolor[HTML]{548235}\textbf{VH} & \multirow{2}{*}{\begin{tabular}[c]{@{}c@{}}- Gain of time to detect/ \\ respond to VAPT attempt  \end{tabular}}
\\ 
Response & \cellcolor[HTML]{548235}\textbf{VH} & \\ 
Prevention & \cellcolor[HTML]{A9D08E} \textbf{H} & - Improve deterrence \\ \bottomrule
\end{tabular}%
}
\end{centering}
\end{table*}

\section{Conclusion \& Future Work} \label{sect:conc}

MDOs aim to integrate capabilities across all active domains of warfare to achieve coordinated, cross-domain effects against targeted systems. Within this context, Smart Cities and in particular CAVs emerge as critical and highly vulnerable assets, especially to cyberattacks leveraging Black-Box AML techniques. A perfect targets in such attacks can be ML-based IDSs employed for securing CAN-based IVNs. These threats are particularly concerning in the broader context of MDTs and, more specifically, within MSTs. Despite increasing attention, current research on Black-Box AML attacks targeting CAN frame detection remains sparse and at an early stage of development.

This paper evaluates (RQ1) the influence of hyperparameters related to DT-based state-of-the-art TT ensemble models (i.e. RF, GB, XGB), underlying an IDS targeted by the ZOO attack (seen as a SST) in a pure Black-Box scenario, on adversarial example generation time. Additionally, it understand (RQ2) whether the combo of the correct setting of the hyperparameters (under exam) and the AT needed time can slow down or reduce the attacker’s "window of
opportunity". Finally, it qualitatively assess (RQ3) the educational impact on MDTFs of this analysis mapped into the CSS-MDO framework contributing to the integration of artificial agents within MDTFs leading to H-AI MDTFs. The experimental results indicate a direct proportional relationship between the number of bagging trees in RF and boosting rounds in GB with the time required to exploit the ZOO attack, a trend not observed in XGB. Thus, only RF and GB exhibit hyperparameters that may serve as intrinsic defense mechanisms contributing to D-TAI. RF is recommended for its superior robustness considering the reduced AT time needed. Generally, the educational impact on MDTFs of such evidence is rated very high considering the important possibility of controlling the attack timing. All this, leads to labeling the collaboration between human and AI in H-AI MDTFs as functional to increase the resilience of social cyber attacks.

Some future directions of this work: to perform the empirical analysis by considering different values related to the attack hyperparameters (even the default ones), to base the analysis on additional Black-Box (and White-Box) attacks as well as additional state-of-the-art datasets in the automotive context. In addition, it is also consedered to extend the analysis on datasets and IDSs exploited in additional systems of national interest (i.e. IoT networks, aircraft, underwater vehicles). Accompanying the analysis of AT times, it is useful to understand the attack times following the application of the countermeasure. Regarding the educational impact analysis, a clear development is to assess the attack impact also considering a victim TSRS in a MST.

\section{Acknowledgments}
\label{sect:acks}
This work was partially supported by the following projects: SERICS - ''Security and Rights In the CyberSpace - SERICS'' (PE00000014) under the MUR National Recovery and Resilience Plan funded by the European Union - NextGenerationEU; Cyber-Predicy, Avviso “Reti - Sostegno alla ricerca collaborativa”, - Codice Progetto DUQVKW0.

\section*{Declaration on Generative AI}
  The author(s) have not employed any Generative AI tools.

\bibliography{sample-ceur}

\end{document}